\documentclass{ifacconf}
\usepackage{amsmath,amsfonts,amssymb,esvect}
\usepackage{graphicx}      
\usepackage{natbib}        
\usepackage{enumitem}
\usepackage[dvipsnames]{xcolor}
\usepackage{flushend}
\usepackage{epstopdf}
\graphicspath{{Figs/}}
\newtheorem{definition}{Definition}
\newtheorem{assumption}{Assumption}
\newtheorem{remark}{Remark}
\definecolor{myred}{rgb}{0.8,0.1,0.16}
\begin{document}
\begin{frontmatter}

\title{Extremum Seeking Control for Antenna Pointing via Symmetric Product Approximation\thanksref{footnoteinfo}} 

\thanks[footnoteinfo]{This paper has been accepted for presentation at The 2025 Modeling, Estimation and Control Conference (MECC 2025). This is the full version of the paper, which contains complete proofs and additional details not included in the conference version. This work was supported in part by GSoE at CCNY, and in part by the PSC-CUNY Award, jointly funded by The Professional Staff Congress and The City University of New York.}

\author[CCNY]{Bo Wang} 
\author[VU]{Hashem Ashrafiuon} 
\author[VU]{Sergey G. Nersesov} 

\address[CCNY]{Department of Mechanical Engineering, The City College of New York, New York, NY 10031, USA (e-mail: bwang1@ccny.cuny.edu).}
\address[VU]{Department of Mechanical Engineering, Villanova University, Villanova, PA 19085, USA (e-mail: \{hashem.ashrafiuon, sergey.nersesov\}@villanova.edu)}

\begin{abstract}              
This paper investigates extremum seeking control for a torque-controlled antenna pointing system without direct angular measurements. We consider a two-degree-of-freedom (2-DOF) antenna system that receives an unknown signal from its environment, where the signal strength varies with the antenna’s orientation. It is assumed that only real-time measurements of the signal are available. We develop an extremum seeking control strategy that enables the antenna to autonomously adjust its direction to maximize the received signal strength based on the symmetric product approximation. Under suitable assumptions on the signal function, we prove local practical uniform asymptotic stability for the closed-loop system.
\end{abstract}


\end{frontmatter}

\section{Introduction}

Satellite communication is one of the fastest-growing fields in science and technology, enabling global information exchange. Antennas play a crucial role in signal transmission and reception, with the quality of the received signal depending on the satellite's relative position and the antenna's orientation. However, manual adjustments to the antenna's orientation are often impractical, particularly when antennas are mounted on mobile vehicles or used in large-scale systems. Therefore, to ensure the strongest, best-quality signal reception, automatic antenna pointing control systems are essential for accurately aligning antennas with the signal source \citep{mulla2016overview}. 

Traditional antenna pointing control strategies, such as PI, LQG, and $H_\infty$, often assume that the reference attitude of the antenna is known, and the difference between the reference and the antenna attitude as sensed by encoders \citep{gawronski2001antenna,gawronski2007control}. In many practical scenarios, it is reasonable to assume that the antenna attitude is measurable by encoders. However, the reference (desired) attitude of the antenna is usually \textit{unknown}, which is typically determined by the signal source. In general, the desired antenna direction is the one that maximizes the received signal strength. In such cases, traditional antenna pointing control algorithms may not be applied.

Extremum seeking is a real-time model-free optimization approach that is applicable to dynamical systems \citep{krstic2000stability,ariyur2003real,scheinker2017model}. The interest in extremum seeking algorithms is greatly motivated by various real-world problems, which require optimizing the performance of dynamic systems based on measurable but analytically unknown functions. Applications include localization of sources for autonomous vehicles \citep{suttner2019extremum,wang2023underactuated}, maximum power point tracking, and bioreactor growth rate optimization, etc. We refer the reader to \cite{scheinker2024100} for an interesting survey.

According to different types of averaging techniques, extremum seeking schemes can be categorized into classical averaging-based \citep{krstic2000stability,ariyur2003real}, Lie bracket averaging-based \citep{durr2013lie,durr2017extremum}, and symmetric product approximation-based seekers \citep{suttner2022extremum,suttner2023extremum}. The classical averaging-based approach and the Lie bracket averaging-based approach work well for first-order kinematic systems; however, their extension to second-order mechanical systems is limited. The symmetric product approximation-based approach, which is based on the averaging results in \cite{bullo2002averaging} for mechanical systems under vibrational control, is applicable to a much larger class of (kinematic-kinetic) mechanical systems. Applications of this strategy
to autonomous vehicles can be found in \cite{suttner2019extremum,wang2023underactuated}.

Applying extremum seeking strategies to antenna pointing systems is intriguing because, as mentioned earlier, the desired antenna attitude is determined by an analytically unknown signal function and is typically unknown. Extremum seeking methods offer an alternative approach to adjusting the antenna’s orientation to maximize the received signal strength. In 
\cite{shore2024extremum}, three classical averaging-based extremum seeking methods for antenna pointing are compared under different simulation scenarios. The methods proposed in \cite{shore2024extremum} rely on the Taylor series approximation of the objective function, however, no stability analysis or results are provided. 

In this paper, we study the problem of extremum seeking control for a torque-controlled antenna pointing system. We consider a two-degree-of-freedom (2-DOF) antenna system that receives an unknown signal from its environment, where the signal strength varies with the antenna’s orientation. Only real-time signal measurements are assumed to be available. 
We develop an extremum seeking control strategy based on the symmetric product approximation, enabling the antenna to autonomously adjust its direction to maximize the received signal strength. The contributions of this paper can be summarized as follows: (i) From the theoretical viewpoint, we provide a stability proof for the closed-loop system. Specifically, we show that under suitable assumptions on the signal function, the origin of the closed-loop system is locally practically uniformly asymptotically stable. (ii) From the practical viewpoint, the proposed seeking scheme does not require posture measurements, relying only on real-time signal measurements. Its structure is exceptionally simple and easy to implement---the measured signal is multiplied by periodic signals and fed into the torque input.
To the best of the authors' knowledge, this marks the first application of the symmetric product approximation-based extremum seeking approach to antenna pointing systems.

The structure of the remaining paper is as follows: Section \ref{sec:problem} presents problem formulation and preliminaries on practical stability. Section \ref{sec:main} presents the main results, which include the extremum seeking control law, averaging, and stability analysis. Section \ref{sec:simulation} presents simulation results illustrating the practical applications of our theoretical findings. Finally, Section \ref{sec:conclusion} offers concluding remarks.

\section{Problem Statement and Preliminaries}\label{sec:problem}

\textit{Notation:} Let $|\cdot|$ denote the Euclidean norm on $\mathbb{R}^n$. For real matrices $A$, we use the matrix norm $\|A\|:=\sup\{|Ax|: |x|=1\}$, and $I_{n\times n}\in\mathbb{R}^{n\times n}$ the identity matrix. Recall that for any real matrix $A:=[a_{ij}]\in\mathbb{R}^{n\times p}$, $\|A\|\le \sqrt{np}\max_{i,j}|a_{ij}|$. We use the notations $s_\theta:=\sin(\theta)$ and $c_\theta:=\cos(\theta)$ for simplicity. Throughout this article, we omit the arguments of functions when they are clear from the context.

\subsection{Modeling of the Antenna Pointing System} 

\begin{figure}[t]
\begin{center}
\includegraphics[width=6.5cm]{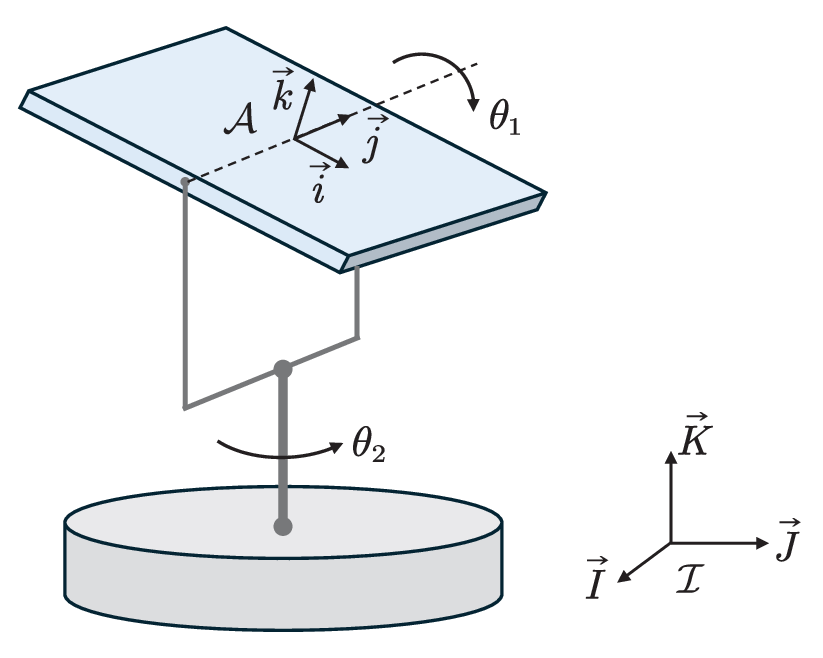}    
\caption{Schematic diagram of the antenna pointing system.} 
\label{fig:antenna}
\end{center}
\end{figure}

Consider a simplified 2-DOF antenna system shown in Fig \ref{fig:antenna}. It is modeled as a rectangular cuboid supported by a frame with respect to which the antenna rotates by the angle $\theta_1$. This frame in turn is attached to the disk that rotates about the vertical axis by the angle $\theta_2$. Both rotations are actuated by two independent motors whose torques are control variables $\tau_1$ and $\tau_2$, respectively. Two reference frames are introduced to describe the motion of the antenna: the stationary reference frame $\mathcal{I}$, with unit vectors $\vv{I}$, $\vv{J}$, and $\vv{K}$, and the body-fixed reference frame $\mathcal{A}$, whose coordinate system, with unit vectors $\vv{i}$, $\vv{j}$, and $\vv{k}$, is fixed at the mass center of the antenna. Since the antenna is modeled as a rectangular cuboid, axes passing through its mass center that are parallel to the unit vectors $\vv{i}$, $\vv{j}$, $\vv{k}$ are principal axes of inertia. In the principal axes of inertia, the equations of rotational motion for the antenna are given by \vspace{-0.3cm}
{\begin{subequations}\label{eqnmotion}
    \begin{eqnarray}
    M_{x}&=&I_x\dot{\omega}_{x}-(I_y-I_z)\omega_y\omega_z,\\
    M_{y}&=&I_y\dot{\omega}_{y}-(I_z-I_x)\omega_z\omega_x,\\
    M_{z}&=&I_z\dot{\omega}_{z}-(I_x-I_y)\omega_x\omega_y,
    \end{eqnarray}
\end{subequations}}
where $I_x$, $I_y$, $I_z$ are antenna's principal moments of inertia; $\omega_{x}$, $\omega_{y}$, $\omega_{z}$ are components (in projection onto $\mathcal{A}$) of the antenna's angular velocity vector $\vv{\omega}_{\mathcal{A}/\mathcal{I}}$ with respect to $\mathcal{I}$, \textit{i.e.}, $\vv{\omega}_{\mathcal{A}/\mathcal{I}}=\omega_x\vv{i}+\omega_{y}\vv{j}+\omega_{z}\vv{k}$. The terms $M_x$, $M_y$, $M_z$ are components (in projection onto $\mathcal{A}$) of the total external moment $\vv{M}$ acting on the antenna with respect to its mass center, \textit{i.e.}, $\vv{M}=M_x\vv{i}+M_y\vv{j}+M_{z}\vv{k}$. 

Next, we relate the angular rates $\omega_{x}$, $\omega_{y}$, $\omega_{z}$ with the angles $\theta_1$ and $\theta_2$. Note that $\vv{\omega}_{\mathcal{A}/\mathcal{I}}=\dot{\theta}_1\vv{j}+\dot{\theta}_{2}\vv{K}$ and $\vv{K}=-\sin\theta_1\vv{i}+\cos\theta_1\vv{k}$. Thus, 
\begin{equation*}
    \vv{\omega}_{\mathcal{A}/\mathcal{I}}=-\dot{\theta}_{2}\sin\theta_1\vv{i}+\dot{\theta}_{1}\vv{j}+\dot{\theta}_{2}\cos\theta_1\vv{k},
\end{equation*}
which implies that 
\begin{equation}\label{omega}
    \omega_x=-\dot{\theta}_{2}\sin\theta_1,\quad
    \omega_y=\dot{\theta}_{1},\quad
    \omega_z=\dot{\theta}_{2}\cos\theta_1.
\end{equation}
The total external moment $\vv{M}$ acting on the antenna consists of two torques rotating it about $\vv{j}$ and $\vv{K}$ axes {and corresponding rotational damping moments along those axes, \textit{i.e.}, $\vv{M}=(\tau_1-d_1\dot{\theta}_{1}) \vv{j} + (\tau_2-d_2\dot{\theta}_{2}) \vv{K}$, where $d_1>0$ and $d_2>0$ are rotational damping coefficients.} Substituting $\vv{K}=-\sin\theta_1\vv{i}+\cos\theta_1\vv{k}$ into $\vv{M}$ yields\vspace{-0.2cm}
\begin{subequations}\label{moments}
\begin{eqnarray}
    M_x&=&-\tau_2\sin\theta_1+d_2\dot{\theta}_{2}\sin\theta_{1},\\
    M_y&=&\tau_{1}-d_1\dot{\theta}_{1},\\
    M_z&=&\tau_2\cos\theta_1-d_2 \dot{\theta}_{2}\cos\theta_1.
\end{eqnarray}
\end{subequations}
Here $\tau_{1}$ and $\tau_{2}$ are control variables. 

Consequently, the model of the antenna pointing system can be in the following Euler-Lagrangian form \vspace{-0.3cm}
\begin{subequations}
  \label{eq:EL}
  \begin{eqnarray}
     \label{eq:EL-a} & \dot{\theta} =J(\theta)\omega &\\
     \label{eq:EL-b} & I\dot{\omega}+C(\omega)\omega + D\omega=J(\theta)^\top\tau  & 
  \end{eqnarray}
\end{subequations}
where $\theta:=[\theta_1~\theta_2]^\top$, $\omega:=[\omega_x~\omega_y~\omega_z]^\top$, $\tau:=[\tau_1~\tau_2]^\top$, $I:=\operatorname{diag}\{I_x,I_y,I_z\}>0$, {$D:=\operatorname{diag}\{d_2,d_1,d_2\}>0$}, and
\begin{equation*}
    J(\theta):=\begin{bmatrix}
        0 & 1 & 0 \\
        -s_{\theta_1} & 0 & c_{\theta_1}
    \end{bmatrix}, 
    C(\omega):=\begin{bmatrix}
        0 & I_z\omega_z & -I_y\omega_y \\
        -I_z\omega_z & 0  & I_x\omega_x \\
        I_y\omega_y & -I_x\omega_x & 0
    \end{bmatrix}.
\end{equation*}

\subsection{Problem Statement}

We assume that the antenna receives an analytically unknown signal from its environment, and the received power $p$ is measurable in real-time. The objective is to design a feedback control law $\tau$ to adjust the antenna's orientation $\theta$ to maximize the received power $p(\theta(t))$, \textit{i.e.}, to solve the maximization problem $\max_{\tau}p(\theta)$ in real-time. However, to maintain consistency with the extremum-seeking literature and without loss of generality, we consider the equivalent minimization problem, \textit{i.e.}, $\min_{\tau}h(\theta)$, where $h(\theta):=-p(\theta)+p_0$ and $p_0$ can be an arbitrary real number. We make the following assumptions on the objective function $h$.
\begin{assumption}\label{assumption:0}
    Assume that the angle-dependent nonlinear cost function $h(\theta)$ is smooth, positive everywhere, and has a global minimum point. That is, there exists a unique $\theta_*:=[\theta_{1*}~\theta_{2*}]^\top$ such that
    \begin{equation}
        \frac{\partial h}{\partial \theta}(\theta_*)=0,\quad\text{and}\quad\frac{\partial h}{\partial \theta}(\theta)\ne 0,\quad \forall\theta\ne \theta_*.
    \end{equation}
\end{assumption}
\begin{assumption}\label{assumption:1}
    The cost function $h$ is in the separable form
    \begin{equation}
        h(\theta)=h_1(\theta_1)+h_2(\theta_2).
    \end{equation}
\end{assumption}
\begin{assumption}\label{assumption:2}
    The Hessian matrix of the cost function is globally bounded. That is, there exists $h_M>0$ such that 
    \begin{equation}
        \left \| \frac{\partial^2 h}{\partial \theta^2} (\theta)\right\|\le h_M,\quad \forall \theta.
    \end{equation}
\end{assumption}

The antenna can measure the value of $h(\theta(t))$ in real-time. Note that both the
extremum $\theta_*$ and the gradient $\frac{\partial h}{\partial \theta}(\theta)$ are unknown.

\begin{remark}\label{rmk:1}
    Assumption \ref{assumption:0} is a regular assumption in the extremum-seeking literature. Assumptions \ref{assumption:1} and \ref{assumption:2} are technical assumptions. Specifically, Assumption \ref{assumption:1} shows that the partial derivatives of $h$ in each direction are independent. One should notice that the objective function used in \cite{shore2024extremum} satisfies Assumptions \ref{assumption:0}-\ref{assumption:2}. In fact, modeling the received power as a Gaussian function of the pointing error yields a quadratic objective function that satisfies Assumptions \ref{assumption:0}-\ref{assumption:2}.
\end{remark}

\subsection{Practical Stability}

We use the same stability notion as in \cite{moreau2000practical,durr2013lie}. Consider the time-varying system that depends on a parameter $\varepsilon\in(0,\varepsilon_0]$
\begin{equation}\label{eq:nltv}
    \dot{x}=f^\varepsilon (t,x),\quad x\in\mathbb{R}^n,
\end{equation}
where $\varepsilon_0>0$. We assume that for each $ \varepsilon $, the function $ f^\varepsilon $ is continuous, and $ f^\varepsilon(t,\cdot) $ is locally Lipschitz uniformly in $ t $. For every $\varepsilon\in(0,\varepsilon_0]$ and every $t\in\mathbb{R}$, let $\phi_t^\varepsilon:\mathbb{R}^n\to \mathbb{R}^n$ be a diffeomorphism. Denote the new variable as $\tilde{x}:=\phi_t^\varepsilon(x)$.

\begin{definition}[\cite{durr2013lie}]\label{def:1} Let $x_*\in\mathbb{R}^n$ and let $S$ be a neighborhood of $x_*$ in $\mathbb{R}^n$. We say that $x_*$ is  \textit{$S$-practically uniformly asymptotically stable} for \eqref{eq:nltv} in the variable $\tilde{x}:=\phi_t^\varepsilon(x)$ if the following two conditions are both satisfied:
    \begin{enumerate}[label=(\roman*)]
        \item \textit{Practical Uniform Stability.} For every $c_1>0$, there exist $c_2$ and $\varepsilon_0>0$ such that for all $\varepsilon\in(0,\varepsilon_0]$ and all $t_0\in\mathbb{R}$
        \begin{equation*}
            |\tilde{x}_0 - x_*|\le c_2 \implies |\tilde{x}(t)-x_*|\le c_1,\quad \forall~t\ge t_0,
        \end{equation*}
        where $\tilde{x}(t):=\phi_t^\varepsilon(x(t))$ and $x(t)$ represents the solution of \eqref{eq:nltv} with initial condition $x(t_0)=\left( \phi_t^\varepsilon\right)^{-1}(\tilde{x}_0)$.
        
        \item \textit{$S$-Practical Uniform Attractivity.} For all $c_1,c_2>0$, there exist $T$, $R$, and $\varepsilon_0>0$ such that for all $\varepsilon\in(0,\varepsilon_0]$, all $t_0\in\mathbb{R}$, and all $\tilde{x}_0 \in S$ with $|\tilde{x}_0 - x_*|\le c_1$
        \begin{equation*}
            \begin{aligned}
                |\tilde{x}(t)-x_*|\le R,\quad &\forall~t\ge t_0,\\
                |\tilde{x}(t)-x_*|\le c_2,\quad &\forall~t\ge t_0 + T,
            \end{aligned}
        \end{equation*}
        where $\tilde{x}(t):=\phi_t^\varepsilon(x(t))$ and $x(t)$ represents the solution of \eqref{eq:nltv} with initial condition $x(t_0)=\left( \phi_t^\varepsilon\right)^{-1}(\tilde{x}_0)$.
    \end{enumerate}
\end{definition}

If the conditions in Definition \ref{def:1} are satisfied for an unknown, arbitrarily small neighborhood $S$ of $x_*$ in $\mathbb{R}^n$, then we say that $x_*$ is \textit{locally} practically uniformly asymptotically stable for \eqref{eq:nltv} in the variable $\tilde{x}$. If there exists a vector field $\bar{f}$ such that $f^\varepsilon=\bar{f}$ for all $\varepsilon\in(0,\varepsilon_0]$, then we omit the word ``practical" in Definition \ref{def:1}.

Next, let us consider the time-varying system
\begin{equation}\label{eq:nltv2}
    \dot{\bar{x}}=\bar{f}(t,\bar{x}),\quad \bar{x}\in\mathbb{R}^n,
\end{equation}
where $\bar{f}$ is locally Lipschitz in $x$ uniformly in $t$.

\begin{definition}[Converging Trajectories Property]
    We say that \textit{the solutions of \eqref{eq:nltv} in the variable $\tilde{x}:=\phi_t^\varepsilon(x)$ approximate the solutions of \eqref{eq:nltv2}}, if for every $T,d>0$ and every compact set $K\subset\mathbb{R}^n$, there exists $\varepsilon_0>0$ such that for all $t_0\in\mathbb{R}$, and all $\bar{x}_0\in K$, the following implication holds: If the solution of \eqref{eq:nltv2} with initial condition $\bar{x}(t_0)=\bar{x}_0$ satisfies $\bar{x}(t)\in K$ for every $t\in[t_0,t_0+T]$, then for all $\varepsilon\in(0,\varepsilon_0]$, we have $|\tilde{x}(t)-\bar{x}(t)|\le d$, $\forall~t\in[t_0,t_0+T]$, where $\tilde{x}(t):=\phi_t^\varepsilon(x(t))$ and $x(t)$ represents the solution of \eqref{eq:nltv} with initial condition $x(t_0)=\left( \phi_t^\varepsilon\right)^{-1}(\tilde{x}_0)$.
   
\end{definition}

\begin{prop}[\cite{suttner2023extremum}]\label{prop:1}
    Let $x_*\in\mathbb{R}^n$ and let $S$ be a neighborhood of $x_*$ in $\mathbb{R}^n$. Assume that the solutions of \eqref{eq:nltv} in the variable $\tilde{x}:=\phi_t^\varepsilon(x)$ approximate the solutions of \eqref{eq:nltv2}. If $x_*$ is $S$-uniformly asymptotically stable for \eqref{eq:nltv2}, then $x_*$ is $S$-practically uniformly asymptotically stable for \eqref{eq:nltv} in the variable $\tilde{x}$.
\end{prop}

\section{Main results}\label{sec:main}

\subsection{Control Law}
Now we introduce the component of the control law. Let $T$ be a positive real number. Let $u_1,u_2:\mathbb{R}\to\mathbb{R}$ be two $T$-periodic, zero-mean, independent perturbation signals, where the zero-mean antiderivatives $U_1$ and $U_2$ of $u_1$ and $u_2$ satisfy the orthogonality condition
\begin{equation}
    \int_0^T U_i(\tau) U_j(\tau) {\rm d}\tau=\left\{\begin{aligned}
        \tfrac{T}{2} & {~~\text{if}~i=j,}\\
         0  & {~~\text{if}~i\neq j}
    \end{aligned}  \right.
\end{equation}
for $i,j\in\{1,2\}$. For instance, we can choose $u_1$ and $u_2$ as $\cos(q t)$ and $\sin(q t)$ for any $q\neq 0$. The proposed extremum seeking control law is given by
\begin{subequations}\label{eq:control}
    \begin{eqnarray}
         \tau_1&=& \frac{k_1}{\varepsilon}u_1\left(\frac{t}{\varepsilon}\right)h(\theta),\\
         \tau_2&=& \frac{k_2}{\varepsilon}u_2\left(\frac{t}{\varepsilon}\right)h(\theta),
    \end{eqnarray}
\end{subequations}
where $\varepsilon$, $k_1$, and $k_2$ are positive control parameters. With the control law \eqref{eq:control} we obtain the closed-loop system
\begin{subequations}
  \label{eq:closed-loop}
  \begin{eqnarray}
     \label{eq:closed-loop-a}  \dot{\theta} &=&J(\theta)\omega \\
     \label{eq:closed-loop-b}  I\dot{\omega}+C(\omega)\omega + D\omega &=& J(\theta)^\top \left[\frac{h(\theta)}{\varepsilon} K u\left(\frac{t}{\varepsilon}\right)\right],  
  \end{eqnarray}
\end{subequations}
where $u(\cdot):=[u_1(\cdot)~u_2(\cdot)]^\top$ and $K:=\operatorname{diag}\{k_1,k_2\}$. Next, our averaging analysis will show that the term in the bracket of \eqref{eq:closed-loop-b} approximates the negative gradient of the function $ \frac{1}{4}h^2 $.

\subsection{Averaging Analysis}

Our goal is to derive the averaged system from the closed-loop system \eqref{eq:closed-loop} using the symmetric product approximation. Here, we recall the approach proposed in \cite{bullo2002averaging,bullo2019geometric} and represent the results \textit{in coordinates}. Let us rewrite the closed-loop system \eqref{eq:closed-loop-b} in the control-affine form
\begin{equation}
    \dot{\omega}=Y_0(\omega)
    +\sum_{i=1}^2 \frac{1}{\varepsilon}u_i\left(\frac{t}{\varepsilon}\right)Y_i(\theta),
\end{equation}
where
\begin{equation*}
    Y_0(\omega):=-I^{-1}C(\omega)\omega-I^{-1}D\omega,
\end{equation*}
\begin{equation*}
    Y_1(\theta):=\begin{bmatrix}
        0\\ I_y^{-1}k_1h(\theta)\\ 0
    \end{bmatrix},\quad
    Y_2(\theta):=\begin{bmatrix}
        -I_x^{-1}s_{\theta_1} k_2h(\theta)\\ 0\\ I_z^{-1}c_{\theta_1}k_2h(\theta)
    \end{bmatrix}.
\end{equation*}
Then, it follows from \cite{bullo2002averaging,wang2023underactuated} that the closed-loop system  \eqref{eq:closed-loop} can be approximated by the \textit{symmetric product system}
\begin{subequations}
  \label{eq:AEL}
  \begin{eqnarray}
       \dot{\bar{\theta}} &=&J(\bar{\theta})\bar{\omega}  \label{eq:AEL-a}\\
       I\dot{\bar{\omega}}+C(\bar{\omega})\bar{\omega} + D\bar{\omega}&=&-\frac{I}{4} \left[\langle Y_1{\,:\,}Y_1\rangle(\bar{\theta}) + \langle Y_2{\,:\,} Y_2\rangle(\bar{\theta}) \right]\quad\quad\label{eq:AEL-b}
  \end{eqnarray}
\end{subequations}
where the \textit{symmetric product} of two vector fields $X,Y\colon \mathbb{R}^2\to \mathbb{R}^3$ corresponding
to the system \eqref{eq:EL} is given by
\begin{equation}
    \langle X{\,:\,}Y \rangle(\theta) =\frac{\partial X}{\partial \theta}(\theta)J(\theta)Y(\theta) + \frac{\partial Y}{\partial \theta}(\theta)J(\theta)X(\theta).
\end{equation}
In \eqref{eq:AEL}, the symmetric products $\langle Y_i{\,:\,}Y_i\rangle$ originates from an iterated
Lie bracket on the tangent bundle of the configuration manifold. We refer the readers to the proof of \cite[Theorem 2]{wang2023underactuated} for more details. It follows from \cite{bullo2002averaging,wang2023underactuated} that the solutions of the closed-loop system \eqref{eq:closed-loop} in the variables
\begin{equation}\label{eq:variable}
    \left(\tilde{\theta},\tilde{\omega}\right):=\left(\theta,\omega-\sum_{i=1}^2 U_i\left(\frac{t}{\varepsilon}\right)Y_i(\theta) \right)
\end{equation}
approximate the solutions of \eqref{eq:AEL}. A direct computation shows that
\begin{equation}\label{eq:Y1}
    \langle Y_1{\,:\,}Y_1 \rangle (\theta) = 2 k_1^2 I_y^{-2}  \frac{\partial h}{\partial \theta_1}(\theta) h(\theta) \begin{bmatrix}
        0 \\ 1 \\ 0
    \end{bmatrix},
\end{equation}
\begin{equation}\label{eq:Y2}
    \langle Y_2{\,:\,}Y_2 \rangle (\theta)=2k_2^2r(\theta_1) \frac{\partial h}{\partial \theta_2} (\theta)h(\theta)\begin{bmatrix}
       -I_x^{-1}s_{\theta_1} \\ 0 \\I_z^{-1}c_{\theta_1}
    \end{bmatrix},
\end{equation}
where $r(\theta_1):=I_x^{-1}s_{\theta_1}^2+I_z^{-1}c_{\theta_1}^2$. Substituting \eqref{eq:Y1}-\eqref{eq:Y2} into \eqref{eq:AEL} yields the \textit{averaged system}
\begin{subequations}
  \label{eq:averaged}
  \begin{eqnarray}
       \dot{\bar{\theta}} &=&J(\bar{\theta})\bar{\omega}  \label{eq:averaged-a}\\
       I\dot{\bar{\omega}}+C(\bar{\omega})\bar{\omega} + D\bar{\omega}&=& J(\bar{\theta})^\top \Lambda(\bar{\theta}_1)\left[-\frac{1}{2}\frac{\partial h}{\partial \theta}^\top(\bar{\theta}) h(\bar{\theta})\right] \quad\quad\label{eq:averaged-b}
  \end{eqnarray}
\end{subequations}
where
\begin{equation}
    \Lambda(\bar{\theta}_1):=\begin{bmatrix}
        k_1^2I_y^{-1} &  0 \\
         0 &  k_2^2 r(\bar{\theta}_1)
    \end{bmatrix},
\end{equation}
and the term in the bracket of \eqref{eq:averaged-b} is exactly the negative gradient of the
function $\frac{1}{4}h(\bar{\theta})^2$. We conclude that the solutions of the closed-loop system \eqref{eq:closed-loop} in the variables \eqref{eq:variable} approximate the solutions of \eqref{eq:averaged}.

\subsection{Stability Analysis}

Next, we provide sufficient conditions for the stability of the averaged system \eqref{eq:averaged}. In general, analyzing the stability of \eqref{eq:averaged} directly using Lyapunov methods is challenging due to the presence of the term $\Lambda(\bar{\theta}_1)$. However, the analysis becomes significantly easier if $\Lambda(\cdot)$ is a constant matrix. Notably, the averaged system \eqref{eq:averaged} is reminiscent of \textit{gradient systems} if $\Lambda(\cdot)$ is a constant matrix. Furthermore, if the function $h$ is quadratic, then the averaged system \eqref{eq:averaged} is reminiscent of a Lagrangian system under PD control, a topic that has been well studied since \cite{takegaki1981new}\footnote{We refer the reader to Section 3.1 in \cite{ortega1998euler}.}. In fact, one should notice that the matrix $\Lambda(\cdot)$ is uniformly bounded, and the term $r(\bar{\theta}_1)$ satisfies $r(\bar{\theta}_1)\in[I_z^{-1}, I_x^{-1}]$\footnote{For the antenna panel shown in Fig. \ref{fig:antenna}, $I_x$ is smaller than $I_z$.}. Hence, in the following, we first analyze the averaged system \eqref{eq:averaged} under the assumption that the matrix \( \Lambda(\cdot) \) is constant, \textit{i.e.}, \( \Lambda(\cdot) \equiv \bar{\Lambda}:= \operatorname{diag}\{\lambda_1, \lambda_2\} \), and provide a strict Lyapunov function for the ``frozen dynamics". Then, we extend the analysis to the case where \( \Lambda(\cdot) \) varies, using the time-scale separation method\footnote{We refer the reader to Section 9.6 in \cite{Khalil:1173048}.}.

First, we show asymptotic stability for the corresponding ``frozen dynamics"
\begin{subequations}
  \label{eq:frozen}
  \begin{eqnarray}
       \dot{\bar{\theta}} &=&J(\bar{\theta})\bar{\omega}  \label{eq:frozen-a}\\
       I\dot{\bar{\omega}}+C(\bar{\omega})\bar{\omega} + D\bar{\omega}&=& J(\bar{\theta})^\top \bar{\Lambda}\left[-\frac{1}{2}\frac{\partial h}{\partial \theta}^\top(\bar{\theta}) h(\bar{\theta})\right], \quad\quad\label{eq:frozen-b}
  \end{eqnarray}
\end{subequations}
where $\bar{\Lambda} := \operatorname{diag}\{\lambda_1, \lambda_2\}>0$. Define the new objective function as
\begin{equation}
    \bar{h}(\theta):=\lambda_1h_1(\theta_1)+\lambda_2h_2(\theta_2),
\end{equation}
where functions $h_1$ and $h_2$ are given in Assumption \ref{assumption:1}. Obviously, {the functions 
$h$ and $\bar{h}$ attain their minimum values} at the same point $\theta_*$ due to the fact that $\lambda_1,\lambda_2>0$.

Let us consider the function
\begin{equation}
	V_1(\bar{\theta}-\theta_*,\bar{\omega}):=\frac{1}{2}\bar{\omega}^\top I \bar{\omega} + \frac{1}{4}\bar{h}(\bar{\theta})^2 -  \frac{1}{4}\bar{h}(\theta_*)^2.
\end{equation}
It is clear that the function $V_1$ is position definite. Taking the total derivative along trajectories of \eqref{eq:frozen} yields
\begin{align}
	\dot{V}_1=\;&\bar{\omega}^\top\left[-C(\bar{\omega})\bar{\omega}-D\bar{\omega}+J(\bar{\theta})^\top \bar{\Lambda}\left(-\frac{1}{2}\frac{\partial h}{\partial \theta}^\top(\bar{\theta}) h(\bar{\theta})\right)\right] \notag \\
	& +\frac{1}{2}\bar{h}(\bar{\theta})\frac{\partial \bar{h}}{\partial \theta}(\bar{\theta})J(\bar{\theta})\bar{\omega} \notag\\
	=\;& -\bar{\omega}^\top D \bar{\omega}\le 0, 
\end{align} 
 where we use the property that the matrix $C(\cdot)$ is skew-symmetric and the fact that $\frac{\partial \bar{h}}{\partial \theta}(\theta)=\bar{\Lambda}\frac{\partial h}{\partial \theta}(\theta)$ everywhere. Thus, we conclude that $V_1$ is a weak Lyapunov function for the system \eqref{eq:frozen}, and asymptotic stability of the equilibrium point $(\bar{\theta},\bar{\omega})=(\theta_*,0)$ comes from by verifying the LaSalle's condition on the set $\{\dot{V}_1=0\}$. If the function $h$ is radially unbounded, we conclude that the equilibrium point $(\bar{\theta},\bar{\omega})=(\theta_*,0)$ is globally asymptotically stable.
 
To proceed with the Lyapunov analysis for the averaged system \eqref{eq:averaged}, we need a \textit{strict} Lyapunov function for \eqref{eq:frozen}. Define $d_m:=\min\{d_x,d_y,d_z\}$, $d_M:=\max\{d_x,d_y,d_z\}$, $I_m:=\min\{I_x,I_y,I_z\}$, and $I_M:=\max\{I_x,I_y,I_z\}$. Let us consider the function
\begin{equation}
	V_2(\bar{\theta},\bar{\omega}):=\frac{\partial \bar{h}}{\partial \theta}(\bar{\theta}) J(\bar{\theta}) I\bar{\omega},
\end{equation}
where its total derivative along trajectories of \eqref{eq:frozen} is
\begin{align}
    \dot{V}_2=\;&\bar{\omega}^\top J(\bar{\theta})^\top \frac{\partial^2 \bar{h}}{\partial \theta^2}(\bar{\theta})J(\bar{\theta})I\bar{\omega} + \frac{\partial \bar{h}}{\partial \theta}(\bar{\theta}) \dot{J}(\bar{\theta}) I\bar{\omega}
    \notag \\
    & -\frac{\partial \bar{h}}{\partial \theta}(\bar{\theta}) J(\bar{\theta}) \left[ C(\bar{\omega})\bar{\omega} + D\bar{\omega} + \frac{h(\bar{\theta})}{2}J(\bar{\theta})^\top \frac{\partial \bar{h}}{\partial \theta}^\top(\bar{\theta})\right] \notag \\
    \le \;&  -\frac{1}{2}    h(\theta_*) \left|\frac{\partial \bar{h}}{\partial \theta}(\bar{\theta}) \right|^2+                   6\bar{h}_MI_M|\bar{\omega}|^2   \notag\\
    & + \sqrt{6}\left[I_M|\bar{\omega}| + 3I_M|\bar{\omega}| + d_M\right]\left|\frac{\partial \bar{h}}{\partial \theta}(\bar{\theta}) \right||\bar{\omega}|,\label{eq:dot_V2}
\end{align}
where we use the facts that $\|J(\cdot)\|\le \sqrt{6}$, $J(\cdot)J(\cdot)^\top = I_{2\times 2}$, $\|\dot{J}(\bar{\theta})\|\le \sqrt{6}|\bar{\omega}|$, $\|C(\omega)\|\le 3I_M|\omega|$, $\|\frac{\partial^2 \bar{h}}{\partial \theta^2}(\cdot)\|\le \bar{h}_M$, and $h(\theta)\ge h(\theta_*)$ everywhere.

Note that $|\bar{\omega}|\le \sqrt{\frac{2}{I_m}}\sqrt{V_1}$. Defining the function $P_1:\mathbb{R}_{\ge 0}\to\mathbb{R}_{\ge 0}$ as
\begin{equation}
    P_1(l):=\sqrt{6} d_M+8\sqrt{3 l}\frac{I_M}{\sqrt{I_m}},
\end{equation}
we have
\begin{equation*}
    \dot{V}_2\le -\frac{1}{2}    h(\theta_*) \left|\frac{\partial \bar{h}}{\partial \theta}(\bar{\theta}) \right|^2+  6\bar{h}_MI_M|\bar{\omega}|^2  + P_1(V_1)\left|\frac{\partial \bar{h}}{\partial \theta}(\bar{\theta}) \right||\bar{\omega}|.
\end{equation*}
It follows from Young's Inequality that
\begin{equation*}
    P_1(V_1)\left|\frac{\partial \bar{h}}{\partial \theta}(\bar{\theta}) \right||\bar{\omega}|\le \frac{h(\theta_*)}{4}\left|\frac{\partial \bar{h}}{\partial \theta}(\bar{\theta}) \right|^2 + \frac{1}{h(\theta_*)}P_1(V_1)^2 |\bar{\omega}|^2.
\end{equation*}
Defining the function $P_2:\mathbb{R}_{\ge 0}\to\mathbb{R}_{\ge 0}$ as
\begin{equation}
    P_2(l):=6\bar{h}_MI_M + \frac{1}{h(\theta_*)}P_1(l)^2
\end{equation}
yields
\begin{equation}
    \dot{V}_2\le -\frac{1}{4}    h(\theta_*) \left|\frac{\partial \bar{h}}{\partial \theta}(\bar{\theta}) \right|^2+ P_2(V_1)|\bar{\omega}|^2.
\end{equation}
Finally, define the function $P_3:\mathbb{R}_{\ge 0}\to\mathbb{R}_{\ge 0}$ as
\begin{equation}
    P_3(l):=\frac{1}{d_m}\int_0^l P_2(m){\rm d}m + P_0 l,
\end{equation}
where
\begin{equation}
    P_0:=\max\left\{\frac{12I_M^2}{I_m},\frac{4\bar{h}^2_M}{\bar{h}(\theta_*)}\lambda_{\min}^{-1}\left[ \frac{\partial^2\bar{h}}{\partial \theta^2}(\theta_*)\right]\right\}.
\end{equation}
We have the following result.
\begin{prop}\label{prop:2}
    Let $\lambda_m,\lambda_M$ be two positive real numbers satisfying $\lambda_m<\lambda_M$. For each $\lambda:=(\lambda_1,\lambda_2)\in[\lambda_m,\lambda_M]\times [\lambda_m,\lambda_M]$, the equilibrium point of the system \eqref{eq:frozen} $(\bar{\theta},\bar{\omega})=(\theta_*,0)$ is asymptotically stable. Furthermore, if the function $h$ is radially unbounded, then the equilibrium point $(\bar{\theta},\bar{\omega})=(\theta_*,0)$ is globally asymptotically stable. Moreover, the function $V_\lambda:\mathbb{R}^2\times\mathbb{R}^3\to\mathbb{R}_{\ge 0}$ defined as
    \begin{equation}
        V_\lambda(\bar{\theta}-\theta_*,\bar{\omega}):=V_2(\bar{\theta},\bar{\omega})+P_3(V_1(\bar{\theta}-\theta_*,\bar{\omega}))
    \end{equation}
    is a strict Lyapunov function for the system \eqref{eq:frozen}.
\end{prop}
\begin{pf}
    First, it is noted that the functions $w_1$ and $w_2$ defined by 
    \begin{equation}
        w_1(\bar{\theta}-\theta_*):= \frac{1}{4}\bar{h}(\bar{\theta})^2 -  \frac{1}{4}\bar{h}(\theta_*)^2
    \end{equation}
    and 
    \begin{equation}
        w_2(\bar{\theta}-\theta_*):=\left|\frac{\partial \bar{h}}{\partial \theta}(\bar{\theta}) \right|^2
    \end{equation}
    are continuous and positive definite. That is, $w_i(0)=0$ and $w_i(s)>0$ for all $s\ne 0$ for $i\in\{1,2\}$. It follows from the Taylor expansion that there exists a neighborhood $S$ of $\theta_*$ such that for all $\bar{\theta}\in S$
    \begin{equation}\label{eq:w1}
        w_1(\bar{\theta}-\theta_*) = \frac{\bar{h}(\theta_*)}{4}(\bar{\theta}-\theta_*)^\top \frac{\partial^2 \bar{h}}{\partial \theta^2}(\theta_*)(\bar{\theta}-\theta_*) + \text{H.O.T.}
    \end{equation}
    and
    \begin{align}\label{eq:w2}
        w_2(\bar{\theta}-\theta_*) &= (\bar{\theta}-\theta_*)^\top \left[\frac{\partial^2 \bar{h}}{\partial \theta^2}(\theta_*)\right]^2(\bar{\theta}-\theta_*) + \text{H.O.T.} \notag \\
        &\le \bar{h}_M^2 |\bar{\theta}-\theta_*|^2 + \text{H.O.T.},
    \end{align}
    where $\text{H.O.T.}$ represents the ``higher order terms".
    Hence, in the neighborhood $S$ we have
    \begin{equation}\label{eq:36}
        \frac{1}{2}w_2(\bar{\theta}-\theta_*)+P_0 w_1(\bar{\theta}-\theta_*)\ge \frac{\bar{h}_M^2}{2}|\bar{\theta}-\theta_*|^2 + \text{H.O.T.}
    \end{equation}
    Note that in \eqref{eq:36} the term $\bar{h}_M^2|\bar{\theta}-\theta_*|^2$ dominates the \text{H.O.T.}, and thus, the right-hand side of \eqref{eq:36} is positive definite for $\bar{\theta}\in S$.
    
    Using Young's Inequality we obtain
    \begin{align}
        |V_2(\bar{\theta},\bar{\omega})|&\le  \sqrt{6} I_M\left| \frac{\partial \bar{h}}{\partial \theta}(\bar{\theta}) \right| |\bar{\omega}| \notag\\
        &\le  \frac{1}{2}w_2(\bar{\theta}-\theta_*) + 3I_M^2 |\bar{\omega}|^2. \label{eq:V2}
    \end{align}
     Consequently, for $(\bar{\theta},\bar{\omega})\in S\times \mathbb{R}^3$ we have
	\begin{align}
		\hspace{-0.1cm}V_\lambda(\bar{\theta}-\theta_*,\bar{\omega})&\ge V_2(\bar{\theta},\bar{\omega}) + P_0 V_1(\bar{\theta}-\theta_*,\bar{\omega}) \notag\\
        &\ge \left[\frac{\bar{h}_M^2}{2}|\bar{\theta}- \theta_*|^2+\text{H.O.T.}\right]  + 3I_M^2 |\bar{\omega}|^2,
	\end{align}
    which shows that $V_\lambda$ is positive definite for $(\bar{\theta},\bar{\omega})\in S\times \mathbb{R}^3$.
    Moreover, direct calculations show that
    \begin{align}
        {\dot{V}_\lambda}|_{\eqref{eq:frozen}}&\le -\frac{1}{4}    h(\theta_*) \left|\frac{\partial \bar{h}}{\partial \theta}(\bar{\theta}) \right|^2 - P_0 d_m |\bar{\omega}|^2 \notag \\
        &\le -\frac{1}{4}    h(\theta_*)w_2(\bar{\theta} - \theta_*)- P_0 d_m |\bar{\omega}|^2, \label{eq:40}
    \end{align}
    which is negative definite. Hence, the function $V_\lambda$ is a strict Lyapunov function, which completes the proof. \qed
\end{pf}

\begin{remark}
    In the case where the function $h$ is quadratic, as mentioned in Remark \ref{rmk:1}, the $\text{H.O.T.}$ term in \eqref{eq:w2} is zero, while the $\text{H.O.T.}$ term in \eqref{eq:w1} is positive definite. It follows directly that $V_\lambda$ serves as a \textit{global} strict Lyapunov function for the system \eqref{eq:frozen}.
\end{remark}

Next, we use the strict Lyapunov function $V_\lambda$ for the frozen dynamics to analyze the full dynamics \eqref{eq:averaged}. First, noting that $r(\theta_1):=I_x^{-1}s_{\theta_1}^2+I_z^{-1}c_{\theta_1}^2$ is bounded uniformly in its initial conditions, \textit{i.e.}, $r(\bar{\theta}_1)\in[I_z^{-1}, I_x^{-1}]$, let us substitute the solution $\bar{\theta}_1(t)$ of \eqref{eq:averaged} into $r(\bar{\theta}_1)$ and denote $\bar{r}(t):=r(\bar{\theta}_1(t))$. Then, $\dot{\bar{r}}(t)=(I_x^{-1}-I_z^{-1})\sin(2\bar{\theta}_1(t))\dot{\bar{\theta}}_1(t)$ is locally Lipschitz due to the fact that the system \eqref{eq:averaged} is smooth and following Gronwall's inequality. Consequently, $\dot{\bar{r}}(t)$ is also bounded. With a slight abuse of notation, defining $\lambda_1:=k_1^2 I_y^{-1}$, $\lambda_2(t):=k_2^2\bar{r}(t)$, and $\bar{\Lambda}(t):=\operatorname{diag}\{\lambda_1,\lambda_2(t)\}$, we rewrite the averaged system \eqref{eq:averaged} as
\begin{subequations}
  \label{eq:slow}
  \begin{eqnarray}
       \dot{\bar{\theta}} &=&J(\bar{\theta})\bar{\omega}  \label{eq:slow-a}\\
       I\dot{\bar{\omega}}+C(\bar{\omega})\bar{\omega} + D\bar{\omega}&=& J(\bar{\theta})^\top \bar{\Lambda}(t)\left[-\frac{1}{2}\frac{\partial h}{\partial \theta}^\top(\bar{\theta}) h(\bar{\theta})\right] \quad\quad\label{eq:slow-b}
  \end{eqnarray}
\end{subequations}
and consider \eqref{eq:slow} as a \textit{slowly time-varying system} with the slowly varing parameter $\lambda_2$ because $\dot{\bar{r}}(t)$ is bounded and $k_2$ can be chosen arbitrarily small in $\dot{\lambda}_2(t)=k_2^2\dot{\bar{r}}(t)$. Now, let us consider the Lyapunov function candidate $V_\lambda$ and calculate its total derivative along trajectories of \eqref{eq:slow}, which yields
\begin{equation}\label{eq:42}
    {\dot{V}_\lambda}|_{\eqref{eq:slow}}= {\dot{V}_\lambda}|_{\eqref{eq:frozen}} + \frac{\partial V_\lambda}{\partial \lambda_2}\dot{\lambda}_2(t).
\end{equation}
Note that in \eqref{eq:42}, \({\dot{V}_\lambda}|_{\eqref{eq:frozen}}\) satisfies \eqref{eq:40} and is, therefore, negative definite with respect to \((\bar{\theta} - \theta_*, \bar{\omega})\). Moreover, \(\dot{\lambda}_2(t)\) can be made arbitrarily small by choosing a sufficiently small control parameter \(k_2\).

\begin{prop}\label{prop:3}
    Consider the system \eqref{eq:slow} and its equilibrium point \((\bar{\theta}, \bar{\omega}) = (\theta_*, 0)\). Assume that \(k_1 > 0\). Then, there exists a positive constant \(\bar{k}_2 > 0\) such that for all \(k_2 \in (0, \bar{k}_2]\), the equilibrium point \((\bar{\theta}, \bar{\omega}) = (\theta_*, 0)\) is uniformly asymptotically stable.
\end{prop}
\begin{pf}
We show that $V_\lambda$ is also a Lyapunov function for \eqref{eq:slow}. Following the proof of Proposition \ref{prop:2}, \( V_\lambda \) is positive definite. It remains to show that the right-hand side of \eqref{eq:42} is negative definite in a neighborhood of $(\theta_*, 0)$. To achieve this, we only need to verify that the term \( \frac{\partial V_\lambda}{\partial \lambda_2} \) has a degree greater than or equal to that of \( {\dot{V}_\lambda}|_{\eqref{eq:frozen}} \) in a neighborhood of the equilibrium point \((\bar{\theta}, \bar{\omega}) = (\theta_*, 0)\)\footnote{For example, consider a function \( V(x):= V_1(x) + k V_2(x) \), where \( V_1 \) and \( V_2 \) are two polynomials, and \( k > 0 \). Assume that \( V_1 \) is negative definite and \( V_2 \) is indefinite with \( V_2(0) = 0 \). If \( \operatorname{deg}(V_2) \ge \operatorname{deg}(V_1) \), then for any compact set \( B \) containing the origin, there exists a positive constant \( \bar{k} > 0 \) such that for each \( k \in (0, \bar{k}] \), the term \( V_1 \) dominates \( k V_2 \) in \( B \), and thus, \( V \) is also a negative definite function in \( B \).
}.

Direct calculation shows that
\begin{equation}
    \frac{\partial V_\lambda}{\partial \lambda_2}=\frac{\partial V_2}{\partial \lambda_2}+\left[ \frac{1}{d_m}P_2(V_1)+P_0\right]\frac{\partial V_1}{\partial \lambda_2}.
\end{equation}
Clearly, \( \frac{\partial V_2}{\partial \lambda_2} \) is of the same degree as \( V_2 \), and thus, according to \eqref{eq:V2}, \( \frac{\partial V_2}{\partial \lambda_2} \) is of the same degree as \( w_2(\cdot) + |\bar{\omega}|^2 \). Next, $\frac{\partial V_1}{\partial \lambda_2}=\frac{\partial w_1}{\partial \lambda_2}$ is of the same degree as $w_1$, which has a degree equal to that of $w_2$, following \eqref{eq:w1}-\eqref{eq:w2}. Consequently, we conclude that \( \frac{\partial V_\lambda}{\partial \lambda_2} \) has a degree greater than or equal to that of \( {\dot{V}_\lambda}|_{\eqref{eq:frozen}} \). Then, one can choose a sufficiently small $k_2$ such that the term \({\dot{V}_\lambda}|_{\eqref{eq:frozen}}\) dominates $\frac{\partial V_\lambda}{\partial \lambda_2}\dot{\lambda}_2(t)$ in
\eqref{eq:42}, which completes the proof.	\qed
\end{pf}

Finally, we can state the following consequence of Propositions \ref{prop:1} and \ref{prop:3}, which is our main result.
\begin{prop}\label{prop:4}
    Consider the Euler-Lagrangian system \eqref{eq:EL} with the control law \eqref{eq:control}. Suppose that the objective function $h$ satisfies Assumptions \ref{assumption:0}-\ref{assumption:2}. Then, there exists $\varepsilon_0>0$ such that for every $\varepsilon\in(0,\varepsilon_0]$ and every $k_1>0$, there exists $\bar{k}_2>0$ such that for all $k_2\in(0,\bar{k}_2]$, the equilibrium point of the closed-loop system \eqref{eq:closed-loop} $(\theta,\omega)=(\theta_*,0)$ is locally practically uniformly asymptotically stable in the variables of \eqref{eq:variable}.
\end{prop} 

\section{Simulation Results}\label{sec:simulation}

\begin{figure}[t]
\begin{center}
\includegraphics[width=6.5cm]{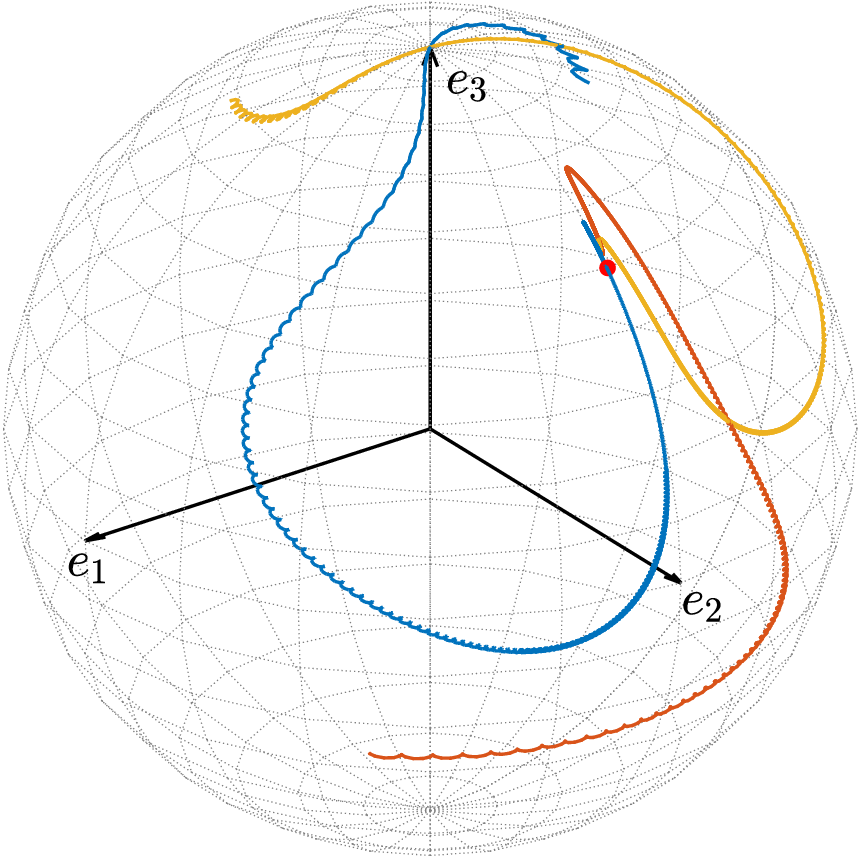}    
\caption{Paths of the attitude of the closed-loop antenna system on $\mathbb{S}^2$, where $(e_1,e_2,e_3)$ denotes the standard basis for $\mathbb{R}^3$. The red dot in the figure represents the desired posture, where the function $h$ attains its minimum value at the point.} 
\label{fig:3D}
\end{center}
\end{figure}

As an example, we consider the antenna system \eqref{eq:EL} with an antenna panel of dimensions: length $= 0.6$ m, width $= 0.3$ m, and height $= 0.1$ m. The mass of the antenna panel is 10 kg.  By calculation, the moments of inertia are determined as $ I_x = 0.0833 $, $ I_y = 0.3083 $, and $ I_z = 0.15 $. The damping coefficients are given by $ d_x = 0.1 $, $ d_y = 0.1 $, and $ d_z = 0.1 $. All parameters are given in SI units. Three initial postures of the antenna are randomly generated as $\theta_{\text{blue}}(0)=[-0.3849,-0.7422]^\top$, $\theta_{\text{red}}(0)=[1.9775,0.7854]^\top$, and $\theta_{\text{yellow}}(0)=[-0.4916, 2.6121]^\top$. All initial angular velocities are assumed to be zero.

In the simulations, we assume that the unknown cost function to be optimized is given by  $h(\theta):= \frac{1}{2}(\theta_1 - \theta_{1d})^2 + \frac{1}{2}(\theta_2 - \theta_{2d})^2 + 0.2$, where \( (\theta_{1d}, \theta_{2d}) = \left( \frac{\pi}{4}, \frac{\pi}{2} \right) \). We choose the input dither signals as \( u_1(t) := \cos(t) \) and \( u_2(t) := \sin(t) \). The control parameters are set to \( \varepsilon = 0.01 \), \( k_1 = 0.2 \), and \( k_2 = 0.1 \). The tracking errors are defined as \( \theta_{1e} := \theta_1 - \theta_{1d} \) and \( \theta_{2e} := \theta_2 - \theta_{2d} \). 

\begin{figure}[t]
\begin{center}
\includegraphics[width=8.4cm]{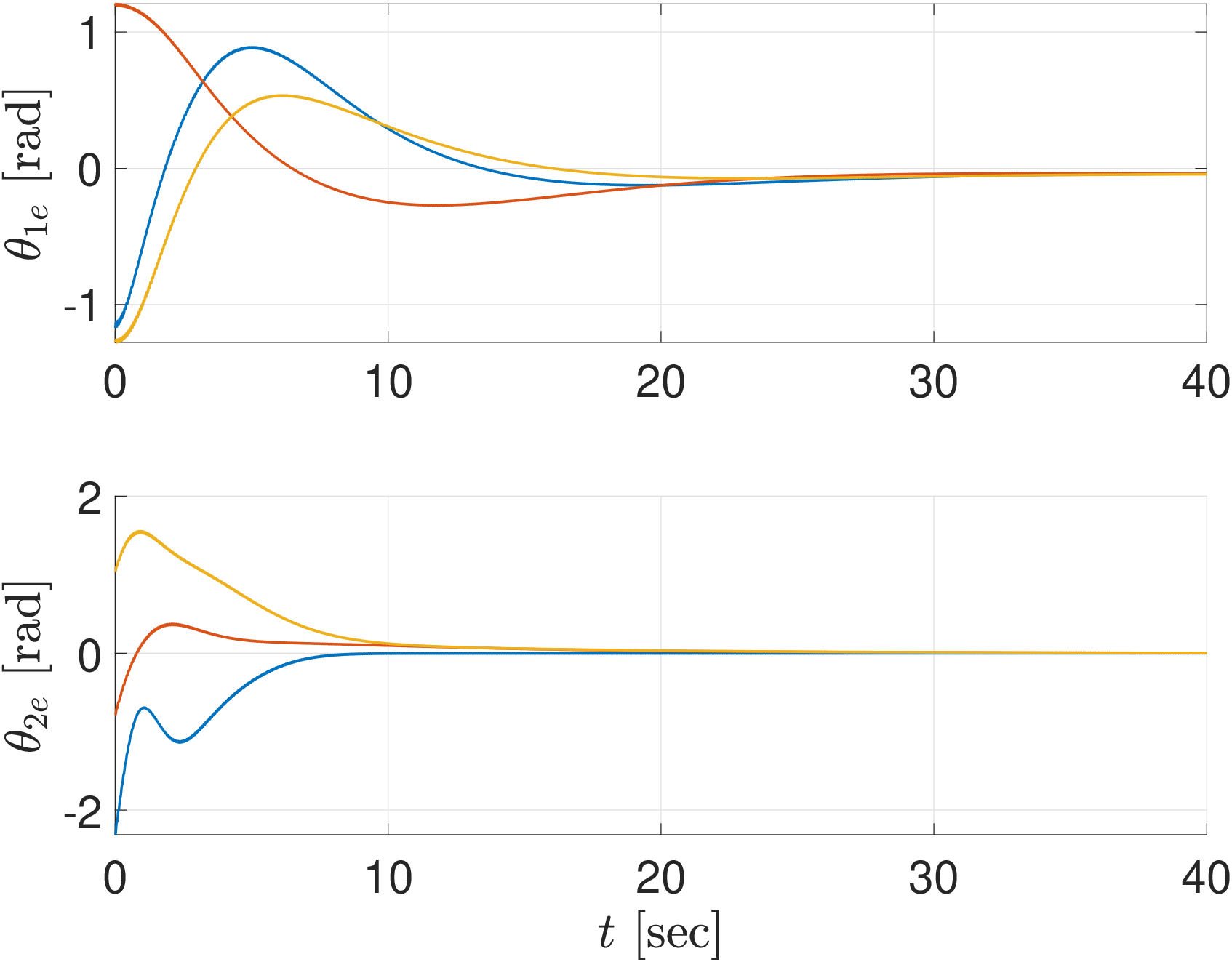}    
\caption{Error trajectories of the antenna pointing system.} 
\label{fig:error}
\end{center}
\end{figure}

The simulation results are presented in Figs. \ref{fig:3D}--\ref{fig:error}. Figure \ref{fig:3D} illustrates the three paths of the antenna system's attitude under the extremum-seeking controller. The red dot in the figure represents the desired posture \( (\theta_{1d}, \theta_{2d}) \), where the function \( h \) attains its minimum value. The figure shows that the antenna’s posture converges to the desired position in all three simulations. Figure \ref{fig:error} displays the error trajectories of the closed-loop antenna pointing system, demonstrating that the tracking errors converge to a small neighborhood around the origin in all three simulations. These simulation results indicate that, although Proposition \ref{prop:4} asserts local practical uniform asymptotic stability, the domain of attraction may not necessarily be small.

\section{Conclusion}\label{sec:conclusion}
In this paper, we study the extremum seeking control problem for a two-degree-of-freedom antenna pointing system, assuming that only real-time signal measurements are available. {The reference (desired) attitude of the antenna is supposed to be unknown.} Utilizing the symmetric product approximation, we propose an extremum seeking control strategy that enables the antenna to adjust its direction to maximize the received signal strength autonomously. Our theoretical analysis demonstrates local practical uniform asymptotic stability for the closed-loop system. Numerical simulations further illustrate the effectiveness of our approach.  {Future research will focus on the development of extremum seeking control strategies for antenna pointing systems subject to time delays, as studied in \cite{oliveira2016extremum,oliveira2022extremum}, along with experimental validation.}

\bibliography{mybibfile}            

\end{document}